\documentstyle[11pt]{article}

\newfont{\ssr}{cmss10 scaled 1100}
\makeatletter
\@addtoreset{equation}{section}

\def\citen#1{%
\edef\@tempa{\@ignspaftercomma,#1, \@end, }
\edef\@tempa{\expandafter\@ignendcommas\@tempa\@end}%
\if@filesw \immediate \write \@auxout {\string \citation {\@tempa}}\fi
\@tempcntb\m@ne \let\@h@ld\relax \def\@citea{}%
\@for \@citeb:=\@tempa\do {\@cmpresscites}%
\@h@ld}
\def\@ignspaftercomma#1, {\ifx\@end#1\@empty\else
   #1,\expandafter\@ignspaftercomma\fi}
\def\@ignendcommas,#1,\@end{#1}
\def\@cmpresscites{%
 \expandafter\let \expandafter\@B@citeB \csname b@\@citeb \endcsname
 \ifx\@B@citeB\relax 
    \@h@ld\@citea\@tempcntb\m@ne{\bf ?}%
    \@warning {Citation `\@citeb ' on page \thepage \space undefined}%
 \else
    \@tempcnta\@tempcntb \advance\@tempcnta\@ne
    \setbox\z@\hbox\bgroup 
    \ifnum0<0\@B@citeB \relax
       \egroup \@tempcntb\@B@citeB \relax
       \else \egroup \@tempcntb\m@ne \fi
    \ifnum\@tempcnta=\@tempcntb 
       \ifx\@h@ld\relax 
          \edef \@h@ld{\@citea\@B@citeB }%
       \else 
          \edef\@h@ld{\hbox{--}\penalty\@highpenalty
            \@B@citeB }%
       \fi
    \else   
       \@h@ld\@citea\@B@citeB
       \let\@h@ld\relax
 \fi\fi%
 \def\@citea{,\penalty\@highpenalty\hskip.13em plus.1em minus.1em}%
}
\def\@citex[#1]#2{\@cite{\citen{#2}}{#1}}%
\def\@cite#1#2{\leavevmode\unskip
  \ifnum\lastpenalty=\z@\penalty\@highpenalty\fi
  \ [{\multiply\@highpenalty 3 #1
      \if@tempswa,\penalty\@highpenalty\ #2\fi 
    }]\spacefactor\@m}
\makeatother

\def\dalemb#1#2{{\vbox{\hrule height .#2pt
        \hbox{\vrule width.#2pt height#1pt \kern#1pt
                \vrule width.#2pt}
        \hrule height.#2pt}}}

\let\a=\alpha    \let\e=\epsilon

 \def\bd{\begin{document}} \def\ed{\end{document}}
\def\ds{\documentstyle} \let\fr=\frac \let\bl=\bigl \let\br=\bigr
\let\Br=\Bigr \let\Bl=\Bigl 
\let\bm=\bibitem
\let\na=\nabla
\let\pa=\partial \let\ov=\overline 
\newcommand{\be}{\begin{equation}} 
\newcommand{\ee}{\end{equation}} 
\def\ba{\begin{array}}
\def\ea{\end{array}}
\def\ft#1#2{{\textstyle{{\scriptstyle #1}\over {\scriptstyle #2}}}}
\def\fft#1#2{{#1 \over #2}}
\def\del{\partial}
\def\sst#1{{\scriptscriptstyle #1}}
\def\oneone{\rlap 1\mkern4mu{\rm l}}
\def\R{\rlap I\mkern3mu{\rm R}}
\def\Z{{\rlap{\ssr Z}\mkern3mu\hbox{\ssr Z}}}
\def\e7{E_{7(+7)}}
\def\td{\tilde}
\def\bog{Bogomol'nyi\ }
\newcommand{\ho}[1]{$\, ^{#1}$}
\newcommand{\hoch}[1]{$\, ^{#1}$}
\newcommand{\bea}{\begin{eqnarray}} 
\newcommand{\eea}{\end{eqnarray}} 
\newcommand{\ra}{\rightarrow}
\newcommand{\lra}{\longrightarrow}
\newcommand{\Lra}{\Leftrightarrow}
\newcommand{\ap}{\alpha^\prime}
\newcommand{\bp}{\tilde \beta^\prime}
\newcommand{\tr}{{\rm tr} }
\newcommand{\Tr}{{\rm Tr} } 
\newcommand{\NP}{Nucl. Phys. }

\begin{document}
\begin{flushright}
\hfill{CTP-TAMU-18/96}\\
\hfill{Imperial/TP/95--96/36}\\
\hfill{hep-th/9605082}\\
\end{flushright}

\vspace{20pt}

\begin{center}
{ \large {\bf Vertical Versus Diagonal Dimensional Reduction for
$p$-branes}}

\vspace{30pt}

H. L\"u\hoch{\dagger}, C.N. Pope\hoch{\dagger}

\vspace{15pt}

{\it Center for Theoretical Physics,
Texas A\&M University, College Station, Texas 77843}
\vspace{10pt}

K.S. Stelle\hoch{\sst\star}

\vspace{15pt}

{\it The Blackett Laboratory, Imperial College, Prince Consort Road, London
SW7 2BZ, UK} 

\vspace{40pt}

\underline{ABSTRACT}
\end{center}

     In addition to the double-dimensional reduction procedure that employs
world-volume Killing symmetries of $p$-brane supergravity solutions and acts
diagonally on a plot of $p$ versus spacetime dimension $D$, there exists a
second procedure of ``vertical'' reduction. This reduces the
transverse-space dimension via an integral that superposes solutions to the
underlying Laplace equation.  We show that vertical reduction is also
closely related to the recently-introduced notion of intersecting
$p$-branes. We illustrate this with examples, and also construct a new
$D=11$ solution describing four intersecting membranes, which preserves
$1/16$ of the supersymmetry.  Given the two reduction schemes plus duality
transformations at special points of the scalar modulus space, one may
relate most of the $p$-brane solutions of relevance to superstring theory.
We argue that the maximum classifying duality symmetry for this purpose is
the Weyl group of the corresponding Cremmer-Julia supergravity symmetry
$E_{r(+r)}$. We also discuss a separate class of duality-invariant
$p$-branes with $p=D-3$. 

{\vfill\leftline{}\vfill
\vskip	10pt
\footnoterule
{\footnotesize
	\hoch{\dagger}	Research supported in part by DOE 
Grant DE-FG05-91-ER40633 \vskip	-12pt}  \vskip	10pt
{\footnotesize 
        \hoch{\sst\star} Research supported in part by the Commission of the 
European Communities under contract SCI*-CT92-0789} }

\pagebreak
\setcounter{page}{1}

\section{Introduction}

     In this paper, we shall explore the two different types of dimensional
reduction that may be used to relate $p$-brane solutions of supergravity
theories in different dimensions. The better-known of these types of
reduction employs Killing symmetries of the $p$-bane solutions in $D$
dimensions to effect a simultaneous reduction on the world-volume and in the
target space, yielding a $(p-1)$-brane in $(D-1)$ dimensions. This is thus
the field-theoretic analogue of the double dimensional reduction procedure
for $p$-brane {\it actions} \cite{dhis}, which is also applicable to
supergravity {\it solutions}, causing ``diagonal'' movement on the $D$
versus $p$ ``brane-scan.'' Aside from a Weyl rescaling of the metric needed
to maintain the conventional Einstein-frame form of the gravitational
action, this diagonal reduction procedure does not change the asymptotic
falloff behaviour of a solution in the directions transverse to the
world-volume. 

     The second type of reduction has been referred to in the literature as
``constructing periodic arrays'' \cite{cg,k,ghl}. This procedure employs the
zero-force property of $p$-brane solutions, which permits multi-center
solutions as well as single-center ones. We shall present, in some more
detail than previously given, just what this second procedure entails. In
particular, it requires the possibility of stacking up a ``deck'' of an
infinite number of $p$-branes in $D$ dimensions, so that the sum of
potentials in a multi-center solution goes over to an integral, thus
changing the asymptotic falloff behaviour in the transverse directions to
that corresponding to a $p$-brane in $D-1$ dimensions. Thus, we shall call
this ``vertical'' dimensional reduction, corresponding to vertical movement
on the brane-scan. In contrast to some impressions gathered from the
literature, we shall show that while the zero-force properties of
supersymmetric $p$-branes indeed allow this type of reduction, there are in
fact many more non-supersymmetric but extremal solutions that also satisfy
the no-force condition. Vertical dimensional reduction can also be applied
to these cases. 

     This stacking-up procedure breaks down for the case of $(D-3)$-branes
in $D$ dimensions, which have conical asymptotic spacetimes in which each
increment to the number of centers in a multi-center solution causes a
portion of the solid angle at transverse infinity to be removed, eventually
leading to a compactification of the transverse spacetime. Thus, the known
examples \cite{wp,bdrgpt,dilatonic} of $(D-2)$-branes seem to reside in a
different category from the other $p$-branes from this perspective. This is
due in part to the fact that they occur in different theories \cite{r10}:
instead of the standard supergravity theories that we consider here, they
require a dilaton potential term, which may be interpreted as the prefactor
for a zero-form field strength \cite{dilatonic}. 

     The vertical reduction procedure involves two stages, namely an
integration over a continuum of charges of the multi-center solution
distributed uniformly over lines, planes or hyperplanes, followed by an
ordinary Kaluza-Klein reduction over the resulting Killing directions. We
show that at the intermediate stage, the metric configuration in the higher
dimension can sometimes be interpreted as a special case of intersecting
$p$-branes.  Thus we see that some $p$-brane solitons that are
vertical-dimensional reductions of $p$-branes in higher dimensions can also
be viewed as intersections of extended objects in the higher dimension.  In
particular, all the multi-scalar and dyonic solutions whose charges are
carried by field strengths derived from the field strengths of the
higher-dimensional theory admit such an interpretation.  In practice, the
higher dimension of greatest interest is $D=11$, in which case these
lower-dimensional $p$-branes can be interpreted as intersections of
M-branes. 

     The two types of dimensional reduction can be combined with duality
transformations in dimensions where the vertical and diagonal reduction
trajectories cross. For this purpose, it is necessary to focus on the
subgroup of duality transformations that map between {\it distinct}
$p$-brane solutions, {\it i.e.}\ the classifying symmetry for $p$-brane
solutions. We shall argue that the maximal relevant duality group here is
the Weyl group of the Cremmer-Julia symmetry $E_{r(+r)}$, where $r=(11-D)$,
as discussed in \cite{weyl}. In the case of vanishing asymptotic values of
the scalar fields, this is the subgroup of the $U$-duality group \cite{ht}
$E_{r(+r)}(\Z)$ that leaves unchanged the scalar moduli of the theory, {\it
i.e.} the asymptotic values of the scalar fields.  In other words, the Weyl
group is the $U$-duality little group of the scalar vacuum. In analogy to
the case of asymptotic Poincar\'e symmetry in general relativity with a flat
asymptotic metric, it seems natural to interpret this as a rigid,
solution-classifying symmetry, instead of as a ``local'' symmetry describing
different forms of identified solutions. If duality is combined with the two
forms of dimensional reduction at special points in the scalar modulus space
in this way, it welds together into one large family many of the $p$-branes
discussed in connection with string theory.\footnote{This family excludes so
far the $(D-2)$ branes, as discussed above. It may be relevant in this
connection that $(D-3)$ branes are dual to supersymmetric instantons, while
$(D-2)$-branes do not seem to have sensible duals.} For example, by
combining vertical and diagonal dimensional reduction with a simple
Weyl-group duality transformation, one may link the string and 5-brane
solutions of $D=10$ supergravity. 

     The interplay between the two types of dimensional reduction and
duality multiplets suggests the question whether there are duality singlet
representations. The standard $p$-brane solutions form non-trivial duality
multiplets. We shall show, however, that at least in the case of
$(D-3)$-branes, there is a class of solutions that are invariant under an
$SL(2,\R)$ duality symmetry. Whether such a construction can be extended to
solutions invariant under full duality groups remains an open problem. 

     A $p$-brane solution of a $D$-dimensional supergravity theory is a 
configuration describing a flat $d=(p+1)$ dimensional Minkowski world 
volume. The $D$-dimensional metric takes the form 
\be
ds^2 = e^{2A} dx^\mu dx^\nu \eta_{\mu\nu} +
e^{2B} dy^m dy^m ,\label{metricform}
\ee
where $A$ and $B$ are functions only of the $y^m$ coordinates of the
$(D-d)$-dimensional transverse space. The relevant part of the
Lagrangian, containing the metric, a dilatonic scalar and an $n$-index
field strength, is 
\be
{\cal L} = eR -\ft12 (\del \phi)^2 -\fft{1}{2n!} e^{a\phi} F^2_n\ .
\label{genlag}
\ee
Here the constant $a$ can be parameterised by
\be
a^2 = \Delta - \fft{2d\td d}{D-2}\ .\label{avalue}
\ee
where $\td d \equiv D-d-2$.  This Lagrangian can give rise to either
elementary $p$-brane solutions with world volume dimension $d=n-1$, or
to solitonic solutions with $d=D-n-1$.   In the former case, $F_n$
carries an electric-type charge; in the latter case, it carries a magnetic
type charge.  Of particular interest are isotropic solutions where the
functions $A$ and $B$ and the dilatonic scalar field $\phi$ depend only on
the variable $r=\sqrt{y^my^m}$.  These solutions take the form
\cite{stainless} 
\bea
ds^2 &=& \Big(1+\fft{k}{r^{\td d}}\Big)^{-\ft{4\td d}{\Delta(D-2)} }
dx^\mu dx^\nu \eta_{\mu\nu} + \Big( 1 + \fft{k}{r^{\td d}} \Big)^{
\ft{4d}{\Delta (D-2)}} dy^m dy^m \ ,\nonumber\\
e^{\phi} &=& \Big(1+ \fft{k}{r^{\td d}} \Big)^{\ft{2a}{\epsilon \Delta}}
\ ,\label{gensol}
\eea
where $\epsilon = 1$ for elementary solutions and $\epsilon = -1$ for 
solitonic solutions.  The solution (\ref{gensol}) is extremal, and has a
mass per unit $p$-volume $m = \ft{\lambda}{ 2\sqrt{\Delta}}$, and from the
field strength $F_n$ one finds a charge $\ft{\lambda}4$, where $\lambda =
-2\td d k /\sqrt{\Delta}$. It describes a $p$-brane with a single center
located at $r=0$.  The value of $\Delta$ for the 4-form field strength in 
$D=11$ is 4, corresponding to $a=0$, which reflects the fact that there 
is no dilaton in the theory.  On descending through the dimensions, the 
possible values of $\Delta$ proliferate.  When $\Delta=4/k$, supersymmetric 
solutions can arise \cite{lp}.  However, the vast majority of allowed values 
correspond to extremal solutions that are not supersymmetric \cite{lp}.

       Note that in this solution, the functions $A$, $B$ and 
the dilaton $\phi$ are linearly related, namely
\be
d A + \td d B = 0\ ,\qquad \phi = - \fft{\epsilon (D-2) a}{\td d} A\ .
\label{ligne}
\ee
It was shown in \cite{stainless} that a single-scalar isotropic $p$-brane in
$D$ dimensions can be double-dimensionally reduced to a single-scalar
$(p-1)$-brane in $(D-1)$ dimensions, with the same value of $\Delta$.  This
is perhaps surprising, since the Kaluza-Klein procedure introduces an
additional scalar field.  The fact that a linear combination of this and the
original scalar of the $D$-dimensional solution vanishes is a non-trivial
consequence of the precise coefficients in the relations (\ref{ligne}).
Indeed, there are other kinds of $p$-brane solution where these relations
are not satisfied. For example, in the dyonic string in $D=6$, the function
$A$ and the dilaton $\phi$ are independent.  As a consequence, the
Kaluza-Klein procedure gives rise to a two-scalar black hole \cite{lpmulti}
in $D=5$, which carries two independent charges.  If either the electric or
the magnetic charge of the dyon vanishes, the corresponding two-scalar black
hole in $D=5$ reduces to a single-scalar one with $\Delta=4$.  On the other
hand, if both charges are equal, it also reduces to a single-scalar black
hole, but with $\Delta=2$.  It was shown in \cite{weyl} that in the context
of maximal supergravity with vanishing scalar moduli, the two-scalar black
holes in $D=5$ form a 270-dimensional representation of the Weyl group of
the U duality group $E_6(\Z)$.  We have seen that some members of this
multiplet oxidise to isotropic dyonic strings in $D=6$.  Others, on the
other hand, oxidise to boosted strings in $D=6$, which carry either electric
or magnetic charge on a 3-form field strength, together with a momentum
associated with the charge carried by the Kaluza-Klein 2-form.  Another
example where the ans\"atze (\ref{ligne}) are relaxed is provided by the
isotropic single-scalar non-extremal black $p$-branes discussed in
\cite{lpx}; these are the general solutions of the equations of motion. 
These involve non-trivial solutions of the Liouville or Toda equations. 
Double-dimensional reduction does not establish relations between solutions
of this type of relaxed ansatz in different dimensions. 

\section{Multi-center solutions} 

     Now, we shall construct multi-center solutions of the 
supergravity theories.  We begin with the elementary case.
The metric ansatz for a multi-center solution is once more given by 
(\ref{metricform}), and again $A$ and $B$ will be taken to be functions
of the transverse  coordinates $y^m$.  The ansatz for the
$n$-form field strength is
\be
F_{m\mu_1\cdots \mu_{n-1}} = \epsilon_{\mu_1\cdots
\mu_{n-1}} \del_m e^{C}\ ,\label{elef}
\ee
where $C$ is taken to be a function of the transverse space coordinates
$y^m$. It is straightforward to obtain the equations of motion
following from the Lagrangian (\ref{genlag}) and to substitute the
ans\"atze for the field strength and the metric.  As in the case of
single-center solutions, one may obtain simple solutions by making a
further ansatz, namely $dA + \td d B=0$. Here, however, we do not
assume isotropicity in the transverse space. In order to obtain
multi-center solutions, we shall first show that the resulting
equations of motion can be cast into a linear form. The remaining
equations of motion are given by 
\bea
\del_m\del_m \phi = -\ft12 a S^2\ ,&& \del_m\del_m A = \ft{\td
d}{2(D-2)} S^2\ , \nonumber\\ 
d(D-2) (\del_m A)^2 + \ft12 \td d(\del_m\phi)^2 &=&\ft12 \td d S^2\ ,
\label{eom1}
\eea
where
\be
S^2= e^{a\phi -2dA} (\del_m e^{C})^2\ ,\label{s2}
\ee
together with an equation of motion for the field strength $F_n$, which we
shall discuss later. The equations (\ref{eom1}) can be solved by taking
$\phi=-a(D-2) A /\td d$, which implies that
\be
S^2 = \Delta (\del_m\phi)^2/a^2\ ,\label{s2'}
\ee
and
\be
\del_m\del_m \phi + \ft{\Delta}{2a} (\del_m\phi)^2 = 0\ .
\ee
This can be rewritten as the linear Laplace equation
\be
\del_m\del_m e^{\ft{\Delta}{2a}\phi} =0\ ,\label{laplace}
\ee
whose point-charge solutions can be superposed to yield multi-center 
solutions. The function $C$ may be found by combining (\ref{s2}) and
(\ref{s2'}), which may be solved by setting $\del_m e^{C} =
(\sqrt\Delta/a)\,\del_m\phi\, e^{-\ft12a\phi +dA}$. It is easy to verify that
this is consistent with the equation of motion for the field strength $F_n$: 
\be
\del_m\del_m C + \del_m C(\del_m C + \td d \del_m B - d \del_m A
+ a\del_m \phi) = 0\ .
\ee
Thus the elementary multi-center $p$-brane solution is given by
\bea
ds^2 &=& \Big( 1 + \sum_\a \fft{k_\a}{|\vec y-\vec y_\a|^{\td d} } 
\Big)^{-\ft{4\td d}{(D-2) \Delta}} dx^\mu dx^\nu \eta_{\mu\nu} +
\Big( 1 + \sum_\a \fft{k_\a}{|\vec y-\vec y_\a|^{\td d} } 
\Big)^{\ft{4d}{(D-2) \Delta}} dy^m dy^m\ ,\nonumber\\
e^{\ft{\Delta}{2a}\phi} &=& 1 + \sum_{\a} \fft{k_\a}{|\vec y-
\vec y_\a|^{\td d}}\ ,\qquad e^C= \ft2{\sqrt\Delta} 
e^{-\ft{\Delta}{2a}\phi}\ .\label{multisol}
\eea
The above is the generic elementary solution for $\td d \ne 0$.  When 
$\td d =0$, the solution (\ref{multisol}) is modified by the
replacement of $|\vec y-\vec y_\a|^{-\td d}$ by $\log |\vec y-\vec
y_\a|$.  In fact when $\td d=0$, one can
also build a modular-invariant $p$-brane solution, which we shall
discuss in section 4.  The electric charge and the mass per unit
$p$-volume for the solution (\ref{multisol}) are given by $Q =
\sum_\a\lambda_\a/4$ and
$m=
\sum_\a
\lambda_\a/(2\sqrt\Delta)$, where $\lambda_\a = -2 \td d
k_\a/\sqrt\Delta$. 

     Given the above explicit form for elementary solutions, it is
now straightforward to construct solitonic solutions. The solitonic
ansatz for the field strength is given by 
\be
F_{m_1\cdots m_n} = -\epsilon_{m_1\cdots m_n p}\, \del_p \sum_\a
\fft{\lambda_\a/\td d}{|\vec y-\vec y_\a|^{\td d}}\ .\label{solf} 
\ee
The solitonic solution can then be straightforwardly constructed, and 
is in fact given by (\ref{multisol}) after the replacement
$\phi\longrightarrow -\phi$.

     Note that we have obtained here generic solutions for all 
values of $\Delta$.  The values of $\Delta$ for supersymmetric $p$-brane 
solutions are given by $\Delta = 4/k$, where $k$ is the number of 
participating field strengths.  Thus one can construct multi-center
solutions  even for non-supersymmetric cases, providing a
counter-example to the  commonly-held belief that the no-force
condition is satisfied only by  supersymmetric solutions.

\section{Vertical dimensional reduction}    

     As we have seen above, dimensional reduction of a
$p$-brane solution to a supergravity  theory can be implemented in two
quite different ways.  The usual process of Kaluza-Klein dimensional
reduction has the effect of lowering the $p$-brane world volume
dimension at the same time. Thus in this double dimensional reduction
procedure, a $p$-brane solution in $D$ dimensions can be reduced on
the $x^\mu$ coordinates to a $(p-1)$-brane solution in $(D-1)$
dimensions. This keeps the dual dimension $\tilde d=D-d-2$ at a
constant value. By contrast, vertical dimensional reduction works
on the transverse coordinates $y^m$, reducing $D$ while preserving the
dimension of the $p$-brane world volume, thus reducing
$\tilde d$ by one but keeping $d$ constant.

     The two dimensional-reduction procedures are
implemented in very different ways.  This difference stems from the
fact that the fields in  the $p$-brane solutions are functions of the
$y^m$ coordinates of the transverse space, but are independent of the
world-volume coordinates $x^\mu$.  Thus, the diagonal process of
Kaluza-Klein double-dimensional reduction, in which the fields
are taken to be independent of one of the $x^\mu$ coordinates, has no 
essential effect on the structure of the solutions. Essentially,
one is reducing on a coordinate on which the solution does not
depend anyway.  (The metric does in fact undergo a conformal rescaling,
but this is only because in Kaluza-Klein dimensional reduction, one
customarily rescales the metric so as to remain  in the Einstein frame
in the lower dimension.) Thus under double dimensional reduction,
all fields in $(D-1)$ dimensions retain the same form as they had in
$D$ dimensions, with $d$ replaced by $(d-1)$. In particular, $\td 
d$ and $\Delta$, and hence the asymptotic behaviour at infinity,
remain unchanged.

      In the vertical dimensional-reduction procedure, it is $\td d$
rather than $d$ that is reduced.  Consequently, as can be seen from
(\ref{gensol}), the asymptotic behaviour of the fields must change. 
This can be achieved because the equations governing $p$-brane
solutions can be cast into a linear form, and
consequently they admit multi-centered solutions in which single-center
solutions are superposed.  If we align the centers of a uniform
continuum of single-centered solutions along some axis in the
transverse space, then the resulting solution acquires a
translational symmetry in that direction and can thus be interpreted
as a solution with one transverse-space dimension fewer.  This is
analogous to the construction of an electrostatic potential in two
spatial dimensions by superposing the solutions for a continuous line
of point charges in three spatial dimensions.  An important
difference, however, is that the multi-centered $p$-brane solutions
have a zero-force condition, which is necessary for static
equilibrium.  It is sometimes argued that supersymmetry of the
solution is responsible for this zero-force condition.  However, as
we have seen in the previous section, static multi-centered
solutions are also possible for non-supersymmetric configurations.  

     In giving the details of the vertical dimensional reduction
procedure, it is  convenient to consider first the case with $\td d\ge
2$.  We take a continuous stack of $p$-branes with uniform charge
density along the coordinate $z=y^{\sst  D-d}$.  It follows
that the sum over discrete charge centers is replaced by an integral,
\be
\sum_\a \fft{k_\a}{|\vec y-\vec y_\a|^{\td d}} \longrightarrow
\int_{-\infty}^\infty \fft{kdz}{(\td r^2 + z^2)^{\ft{\td d}{2}}} =
\fft{\td k}{\td r^{\td d -1}}\ ,\label{integ}
\ee
where $\td k = k\sqrt{\pi} \Gamma(\td d -\ft12)/(2\Gamma(\td d))$, and
$\td r^2 = y^{\td m} y^{\td m} = (y^{1})^2 + \cdots (y^{D-d-1})^2$.  The
metric of the $p$-brane solution then becomes 
\be
ds^2 = \Big(1+\fft{\td k}{\td r^{\td d-1}}\Big)^{-\ft{4\td d}{\Delta(D-2)} } 
dx^\mu dx^\nu \eta_{\mu\nu}   +
\Big( 1 + \fft{\td k}{\td r^{\td d-1}} \Big)^{ \ft{4d}{\Delta
(D-2)}} ( dz^2 + dy^{\td m} dy^{\td m}) \ .\label{gv}
\ee 
Since this solution is now independent of the coordinate $z=y^{D-d}$, one
may  project it into the $(D-1)$-dimensional subspace. The resulting metric
is not yet the usual one (\ref{gensol}) for a $p$-brane in $(D-1)$
dimensions.  The difference, however, is merely accounted for by scaling
with an overall conformal factor. In other words, the $(D-1)$-dimensional
solution is  obtained in a frame that is not the usual Einstein frame.  If
we multiply  the above metric by a conformal factor $(1+ \td k \td r^{-\td d
 +1})^{4d/(\Delta(D-2)(D-3))}$, the solution takes precisely the form
(\ref{gensol}) in $(D-1)$ dimensions, with the {\it same} value of $\Delta$
as for the original solution in $D$ dimensions. Thus if one aligns a
continuum of $p$-branes with uniform charge density along an axis in the
transverse space, the spacetime configuration is effectively reduced to
$(D-1)$ dimensions, since it is now uniform in the direction of this axis.
From the point of view of the remaining coordinates, the solution is
effectively a $p$-brane in $(D-1)$ dimensions.  As in the case of double
dimensional reduction, the value of $\Delta$ is preserved by the vertical
dimensional reduction procedure. This is consistent with the fact that
vertical dimensional reduction preserves the supersymmetry of the solution.
This preservation of supersymmetry occurs because the conditions for
unbroken supersymmetry depend on the algebraic relations between the
functions $A$, $B$, $C$ and $\phi$, and not on the particular single-center
(\ref{gensol}) or multi-center (\ref{multisol}) form of the solution to the
Laplace equation (\ref{laplace}) 

     The process of vertical dimensional reduction described above can be
thought of as a two-stage procedure.  First, by integrating over a continuum
of single-center solutions along a line in the transverse space, the metric
is reduced to the form (\ref{gv}).  As yet, no reduction in the total
spacetime dimension has taken place.  Although the metric is now independent
of $z$, this coordinate cannot simply be included along with the $p$-brane
world-volume coordinates $x^\mu$ to describe a $(p+1)$-brane, since the
functional dependence of its $\td r$-dependent prefactor differs from that
for the world-volume coordinates.  Nevertheless, the coordinate $z$ is a
world-volume-like coordinate, since the solution is independent of $z$.  In
this first stage, the integration over $z$ gives rise to a new Killing
vector, which generates translations along $z$. We can then, as a second
step, perform the standard Kaluza-Klein reduction on the coordinate $z$,
thereby obtaining a $p$-brane in $(D-1)$ dimensions.  In this example, we
considered the case where the charge is uniformly distributed along a line
parameterised by the coordinate $z$.  To perform vertical dimensional
reduction to lower dimensions, we can apply this procedure iteratively, or
we can directly consider the cases where the charges are uniformly
distributed on a plane or a hyperplane, on which the standard Kaluza-Klein
procedure can then be performed. 

     In cases with the initial value $\td d =1$, the integration
(\ref{integ}) becomes divergent. These cases can be handled by putting a
cutoff and renormalising by subtracting an infinite constant; the result for
the analogue of (\ref{integ}) is of the form $(1 + k\log r)$, which is a
solution to the Laplace equation in two dimensions, corresponding to $\td
d=0$. Thus the $(D-4)$-brane in $D$ dimensions reduces to a $(D-4)$-brane in
$(D-1)$ dimensions, {\it i.e.} it becomes a member of the $(D-3)$-brane
diagonal trajectory. This appears to be the last sensible step for vertical
reduction, however, as the $(D-2)$-branes that lie on the next lower
diagonal trajectory exist \cite{wp,bdrgpt,dilatonic} only in a different
class of supergravity theories \cite{r10} that include a dilatonic-scalar
potential. These dilatonic potentials generalise the cosmological term in
the action, and have the effect of ruling out empty flat space as a
solution.\footnote{It is true that such supergravity theories have recently
been given a common formulation together with the standard ones by the
introduction of a rank $D-1$ antisymmetric-tensor gauge field \cite{bdrgpt},
which has no continuous degrees of freedom but which introduces the
coefficient of the dilaton-scalar potential as an integration constant. It
is not clear to us if this will be helpful in pushing vertical dimensional
reduction beyond the $(D-3)$-brane barrier.} 

     The difficulties in proceeding with vertical dimensional reduction from
$(D-3)$-branes to $(D-2)$-branes is mirrored by a special aspect of the
asymptotic spacetimes for $(D-3)$-branes. The transverse asymptotic spaces
of $(D-3)$-branes are not asymptotically flat, but are instead
asymptotically conical, with a deficit angle related to the ADM energy
density of the solution. Although $(D-3)$-branes do obey the zero-force
property, any attempt to stack up enough of them along an axis so as to
generate the transverse translational symmetry required for vertical
reduction runs into the difficulty that the total $D$-dimensional mass
density then exceeds the limit at which the transverse spacetime is ``eaten
up'' by the deficit angle bites taken out of it. The $(D-3)$-branes thus
appear to be the natural endpoints of the process of vertical dimensional
reduction.

\section{Vertical reduction and intersecting $p$-branes}

     In double-dimensional reduction, an isotropic $p$-brane solution in
$(D-1)$ dimensions can often be viewed as an isotropic $(p+1)$-brane in $D$
dimensions.  On the other hand, in vertical-dimensional reduction, an
isotropic $p$-brane in $(D-1)$ dimensions can no longer be isotropically
oxidised into $D$ dimensions (hence the terminology ``stainless'' introduced
in \cite{stainless}), but as we saw in the previous section it can be
viewed as a $p$-brane solution in $D$ dimensions whose charge is uniformly
distributed along the extra coordinate.  Configurations of this type are
closely related to the recently-introduced notion of intersecting $p$-branes
\cite{pt,t,kt,gkt}. To illustrate this, let us consider the two-scalar black
hole \cite{lpmulti} in $D=7$, with two independent electric charges $q_{12}$
and $q_{34}$ carried  by the 2-form field strengths $F^{(12)}$ and
$F^{(34)}$. These field strengths $F^{(ij)}$ arise from dimensional
reduction of the 4-form field strength $F_{\sst{MNPQ}}$ in $D=11$, in which
two of the indices lie in the internal directions $z_i$ and $z_j$.   This
black hole can be dimensionally oxidised to $D=11$, giving rise to two
intersecting membranes in $D=11$ supergravity, or M-theory.  The two
membranes share a common time coordinate, but have two orthogonal spatial
surfaces, namely $(z_1,z_2)$ and $(z_3,z_4)$ respectively. The
11-dimensional metric is given by \cite{t,gkt} 
\bea
ds^2 &=& -H_{12}^{-2/3} H_{34}^{-2/3} dt^2 + H_{12}^{1/3} H_{34}^{-2/3} 
(dz_1^2 + dz_2^2)\nonumber\\ 
&&+ H_{34}^{1/3} H_{12}^{-2/3} (dz_3^2 + dz_4^2) + H_{12}^{1/3} H_{34}^{1/3}
(dy_1^2 + \cdots + dy_6^2) \ ,\label{2mem}
\eea
where $H_{ij} = 1 + q_{ij}/r^{4}$, and $r^2 = y^2_1 + \cdots y_6^2$.  This 
metric can be interpreted as describing two intersecting membranes 
\cite{t,gkt}.
The reason for this is as follows.  If we consider the limit where 
$q_{12}=0$, the metric then describes a multi-center membrane \cite{ds}, with 
world-volume coordinates $(t,z_1,z_2)$, in which the multiple charges are 
distributed uniformly over the $(z_3,z_4)$ plane.  On the other hand, if 
$q_{34}$ is instead set to zero, the metric describes a different 
multi-center membrane with world-volume coordinates $(t,z_3,z_4)$, with its 
charges distributed uniformly over the $(z_1,z_2)$ plane.  If we now 
consider the general case where $q_{12}$ and $q_{34}$ take generic 
non-vanishing values, we see that the solution (\ref{2mem}) interpolates 
between these two limiting cases, and can be interpreted as describing the 
intersection of the two membranes.  Note that {\it a priori}, neither of the 
limiting cases $q_{12}=0$ or $q_{34}=0$ by itself would deserve the 
interpretation as an intersection of membranes, because the $(z_3,z_4)$ or 
$(z_1,z_2)$ planes respectively only acquire an interpretation as spatial 
membrane world-volume coordinates when one moves away from the limiting 
cases.  It is the ability to interpolate between the two limits that 
provides the compelling interpretation as an intersection of membranes.

     Another limiting case that is of interest is when the two charges
$q_{12}$ and $q_{34}$ are equal.  In this case, the the resulting $\Delta=2$
black hole in $D=7$ becomes a G\"uven solution \cite{g} in $D=11$, which was
interpreted as a special case of two intersecting membranes in \cite{pt}. 
Again, if one had only this special example of the G\"uven solution, where
$H_{12}=H_{34}$, the interpretation as intersecting membranes would be
rather obscure, since in this degenerate limit, there is nothing that breaks
the symmetry of the four coordinates $z_1, z_2, z_3$ and $z_4$.   Thus again 
it seems to be important for this interpretation of the G\"uven solution
that there should exist more general solutions where the harmonic functions
$H_{12}$ and $H_{34}$ can be independent. 

     Having established the interpretation of the class of solutions given 
by (\ref{2mem}), it is of interest to note that the two limiting cases 
$q_{12}=0$ or $q_{34}=0$ can be viewed as the first stage of a vertical 
dimensional reduction of the membrane in $D=11$ to a membrane in $D=9$.  Put 
another way, we see that the $q_{12}=0$ or $q_{34}=0$ limits of the 
solutions (\ref{2mem}) are the Kaluza-Klein oxidations of certain membrane 
solutions in $D=9$.  Thus vertical dimensional reduction provides a way of 
interpreting certain lower-dimensional $p$-branes as intersections of 
extended objects in eleven-dimensional M-theory.

       It should be emphasised, however that not all stainless $p$-brane
solitons can be viewed as intersecting branes in $D=11$.  First of all,
the charges of the $p$-brane in the lower dimension must all be carried by
field strengths that are derived from the 4-form in $D=11$ rather than from the
Kaluza-Klein reduction of the metric, since otherwise the oxidation will
generate off-diagonal ``boosts'' in the the $D=11$ metric.  (In fact, these
types of $p$-branes can be viewed as boosted intersecting M-branes 
\cite{kklp}.) Secondly, even if the solution does satisfy this condition, it
is not always possible to interpret it as intersecting $p$-branes in
$D=11$.  For example, the membrane in $D=10$ can be viewed as the
multi-center membrane in $D=11$ with the charges lying along the extra
dimension $z$, 
\be
ds^2 = H^{-2/3}(-dt^2 +
dx^idx^i) + H^{1/3} (dy^m dy^m + dz^2)\ .\label{sm}
\ee
At first sight, one might interpret this in $D=11$ as a special case of a
membrane intersecting a string. However, this solution lacks the essential
feature that we discussed previously, namely that there was a generalisation
to a metric with an independent harmonic function associated with each of
the intersecting extended objects. The construction of such a generalisation
for the metric (\ref{sm}) (which itself can be diagonally reduced to a
single-scalar black hole in $D=8$) would be equivalent to constructing black
holes in $D=8$ with independent charges for two of the three field strengths
$F^{12}$, $F^{13}$ and $F^{23}$.  However, no simple solution of the
standard multi-scalar-type with two independent charges exists
\cite{lpmulti}.  Thus the extra dimension on which the membrane lies cannot
be elevated to a spatial string worldsheet dimension, and hence it seems to
be unnatural to regard the metric (\ref{sm}) as that of a membrane
intersecting a string.  It is worth remarking that there also exists an
isolated two-charge (non-supersymmetric) black hole solution in $D=8$ where
the two charges are equal, corresponding to $\Delta=3$ \cite{lp}. The
oxidation of this solution to $D=11$ again might be viewed as a membrane
intersecting a string.  However, in this case the three internal coordinates
$z_1$, $z_2$ and $z_3$ enter the metric symmetrically.  Since there is no
generic two-charge solution that permits an extrapolation away from this
degenerate configuration, there is again no compelling reason to regard such
a metric configuration as a membrane intersecting a string.  Indeed, this is 
consistent with the fact that there is no fundamental string in $D=11$.

       An analogous analysis can be applied also to the multi-scalar black
hole in $D=5$ with three independent charges $q_{12}, q_{34}$ and $q_{56}$
carried by the 2-form field strengths $F^{12}$, $F^{34}$ and $F^{56}$
\cite{lp}.  Oxidation of this black hole to $D=11$ gives rise to three
\cite{t,gkt} intersecting membranes. If all the charges are equal, the
resulting $\Delta=4/3$ Reissner-Nordstr\o{m} black hole in $D=5$ becomes a
G\"uven solution in $D=11$.  If one of the charges is zero, it becomes two
multi-center intersecting membranes with the remaining two charges uniformly
distributed over a plane.  When two of the charges are zero, it becomes the
standard multi-center membrane with the remaining charge uniformly
distributed over a 4-dimensional hyperplane. 

     We may construct a further example by starting from the multi-scalar 
black
hole in $D=3$ \cite{weyl}, with four independent electric charges $q_{12},
q_{34}, q_{56}$ and $q_{78}$ carried by the 2-form field strengths $F^{12}$,
$F^{34}$, $F^{56}$ and $F^{78}$.  The metric in $D=3$ takes the form $ds^2 =
-dt^2 + H_{12} H_{34} H_{56} ( dy_1^2 + dy_2^2)$,  where $H_{ij} = 1 +
q_{ij} \log{r}$ and $r^2 = y_1^2 + y_2^2$. We find that dimensional
oxidation of this solution to $D=11$ gives rise to four intersecting
membranes, with the metric 
\bea
ds^2 &=& -(H_{12} H_{34} H_{56} H_{78})^{-2/3} dt^2 + H_{12}^{-2/3} (H_{34}
H_{56} H_{78})^{1/3} (dz_1^2 + dz_2^2) \nonumber\\
&&+ H_{34}^{-2/3} (H_{12}
H_{56} H_{78})^{1/3} (dz_3^2 + dz_4^2) +
H_{56}^{-2/3} (H_{12}
H_{34} H_{78})^{1/3} (dz_5^2 + dz_6^2) \\
&&+ H_{78}^{-2/3} (H_{12}
H_{34} H_{56})^{1/3} (dz_7^2 + dz_8^2) + 
(H_{12} H_{34} H_{56} H_{78})^{1/3} (dy_1^2 + dy_2^2)\ ,\nonumber
\eea 
Thus the G\"uven solutions, which have $p=2,4,6$ and preserve $2^{-p/2}$ of
the supersymmetry, can be extended to include $p=8$, preserving $1/16$ of
the supersymmetry.  This case arises when all four charges $q_{12}$,
$q_{34}$, $q_{56}$ and $q_{78}$ are equal, corresponding to a $\Delta =1$
black hole in $D=3$, which again preserves $1/16$ of the supersymmetry
\cite{weyl}.  

     In summary, we have seen that in $D=11$ one can construct up to four
orthogonally intersecting membranes, and that they reduce to black holes in
$D=9,7,5$ and $3$ dimensions respectively.  This is closely related to the
fact that one needs 2-form field strengths with non-overlapping internal
indices, for example $F^{12}, F^{34}$, {\it etc}, in order to be able to
construct the necessary multi-scalar solutions with independent charges
\cite{lpmulti}.  For $N$ intersecting membranes, there are $N$ independent
charges.  If $n$ of the $N$ charges are zero, the solution becomes $(N-n)$
multi-center intersecting membranes whose charge lies uniformly on a
$2n$-dimensional hyperplane, providing the first stage of the
vertical-dimensional reduction of the solution.  Thus we see that some
stainless $p$-brane solitons that are vertical-dimensional reductions of
$p$-brane in higher dimensions can also be viewed as special cases of
intersecting $p$-branes in the higher dimension.  The criterion in general 
for being able to interpret a $p$-brane in a lower dimension as an 
intersection of extended objects in $D=11$ is that there must exist 
corresponding multi-charge $p$-brane generalisations in the lower dimension.
All the examples we discussed above involved elementary solutions, carrying 
electric charges.  A completely analogous discussion applies to solitonic 
solutions, which carry magnetic charges.

\section{Reduction trajectories linked by duality}

     In the vertical dimensional reduction procedure, the rank of the
participating field strength remains the same for elementary
solutions, whilst it is  reduced by 1 in each step of the reduction
for solitonic solutions. Precisely the opposite happens under
double-dimensional, {\it i.e.}\ diagonal, reduction. If one has an n-index
field strength $F_n$ in $D$ dimensions, its Kaluza-Klein dimensional reduction
gives rise to field strengths $G_n$ and $G_{n-1}$ in $(D-1)$ dimensions. Each
of these can give rise to an elementary or a solitonic isotropic $p$-brane in
the lower dimension.  By making use of both the double dimensional and 
vertical dimensional reduction procedures, all four of the isotropic
solutions in $(D-1)$ dimensions can be obtained from the two isotropic
solutions involving $F_n$ in $D$ dimensions. Specifically, the elementary
$(n-3)$-brane using $G_{n-1}$ and the solitonic $(D-n-3)$-brane using $G_n$
can be obtained by double dimensional reduction, whilst the solitonic
$(D-n-2)$-brane using $G_{n-1}$ and the elementary $(n-2)$-brane using $G_n$
can be obtained by vertical dimensional reduction. In summary, consider a
$(D-1)$-dimensional supergravity that is obtained by Kaluza-Klein
dimensional reduction from a $D$-dimensional supergravity. {\it All} of the
isotropic $p$-brane solutions in $(D-1)$ dimensions using field strengths
other than the 2-form arising from the metric can be obtained either by
double dimensional reduction or by vertical dimensional reduction of the
isotropic solutions one dimension higher in $D$ dimensional supergravity. 

     When the asymptotic values of the scalar fields are taken to vanish, the
$p$-brane solutions with a given value of $p$ form representations of the
Weyl group \cite{weyl} of the Cremmer-Julia $E_{r(+r)}$ supergravity
symmetry group, $r=11-D$. The Weyl-group duality transformations in $D$
dimensions rotate pure electric and pure magnetic solutions into each other,
but do not rotate them into dyons (in even dimensions where dyons exist).
Such transformations can now be used on the various results of double and
vertical dimensional reduction when the scalar moduli take special values.
For example, the elementary string and solitonic 5-brane in (D=10) are built
with the same 3-form field strength. Nonetheless, there is not known to date
any symmetry of $D=10$ supergravity that directly changes the string into
the 5-brane. However, upon vertical reduction of the string down to $D=6$,
one finds, for vanishing scalar moduli, a string that lies in a
representation of the Weyl group of $SO(5,5)$, which is the $D=6$
Cremmer-Julia symmetry group. This Weyl group {\it does} now contain a {\it
bone fide} symmetry transformation that dualises the 3-form field strength
and yields the solitonic string that descends from the $D=10$ 5-brane by
diagonal dimensional reduction. Whether this procedure of linking
reduction/oxidation trajectories with dualities actually indicates the
existence of a symmetry in $D=10$ supergravity is open to debate.
Dimensional reduction of supergravity solutions rely on special properties
of those solutions, such as the existence of Killing vectors or obeying a
zero-force condition that allows them to be stacked up. Nonetheless, the
analogous states in superstring theories are widely anticipated to be linked
by non-perturbative duality symmetries, with the $D=10$ string/5-brane
relation being a familiar example. 

     A similar example of a duality linkage combined with the two types of
dimensional reduction concerns the $D=11$ elementary membrane and solitonic
5-brane. In $D=11$, both of these solutions employ the same field $F_4$
field strength. Upon vertical reduction of the membrane to $D=8$, one
obtains a membrane that lies in the same multiplet of $S_2\times S_3$ (the
Weyl group of the $D=8$ Cremmer-Julia group $SL(3,\R)\times SL(2,\R)$) as
the solitonic membrane descending from the solitonic 5-brane in $D=11$. Here
too, the transformation dualises the relevant $F_4$ field strength. By the
same token as in the $D=10$ case, this suggests the existence of a duality
symmetry of $D=11$ supergravity (or at least $M$ theory) that maps the
membrane and the 5-brane into each other. 

     Duality symmetries also exist that link $p$-branes using different
field strengths. For example, consider a $D=10$ elementary string and
elementary membrane. In type IIA supergravity in $D=10$, there is a single
NS-NS sector $F_3$ field strength, which gives rise to the string, and a
single R-R sector $F_4$ field strength, which gives rise to the membrane.
Upon dimensional reduction of the $D=10$ theory to $D=9$, two three forms
arise, one from the $D=10$ $F_3$ and one from the $F_4$. These two $D=9$
three-forms, $(F_3^1,F_3^2)$ form a doublet under the $D=9$ duality Weyl
group $S_2$ \cite{weyl}. The corresponding duality transformation maps the
$D=9$ string arising by vertical reduction from the string in $D=10$ into
the $D=9$ string arising by diagonal reduction from the $D=10$ membrane. If
lower-dimensional symmetry indicates the existence of a higher-dimensional
symmetry, this would suggest string/membrane duality in $D=10$. 

     The solitonic analogue of the previous example maps the vertical
reduction of the solitonic 4-brane in $D=10$ into the diagonal reduction of
the solitonic 5-brane in $D=10$. While these two examples do not reveal
electric/magnetic (or strong/weak coupling) duality (which is not relevant
in $D=9$), they reveal another characteristic of the unifying role of
duality transformations, in that they link solutions using field strengths
arising in the NS-NS and R-R sectors of superstring theory. 

     One may ask why the duality transformations in these examples are taken
from the Weyl group of the Cremmer-Julia symmetry, and not from the full
group. As discussed in \cite{weyl}, when the asymptotic values of the scalar
fields vanish, the Weyl group arises as the $U$-duality little group of the
scalar field vacuum, {\it i.e.}\ the subgroup of the full $E_r(+r)(\Z)$
group that leaves unchanged the asymptotic values of the scalar fields. The
part of $E_r(+r)$ that acts nontrivially on these moduli is akin to the set
of general coordinate transformations that do not leave the asymptotic form
of the metric invariant. The interpretation of such transformations may vary
according to the context. In general relativity, such transformations are
not customarily regarded as generating distinct new solutions, but merely
are seen as giving equivalent forms of known solutions written in new
coordinates. For example, flat space written in Cartesian or polar
coordinates is regarded as being one and the same solution. 

     On the other hand, the little group of the asymptotic metric is
normally treated differently. For asymptotically flat spaces, this is the
Poincar\'e group, whose transformations are {\it not} treated as yielding
members of an equivalence class to be identified. In fact, if one were to
try to insist on a ``local'' interpretation for such asymptotic symmetries
of the spacetime, one would have to restrict attention to the
Poincar\'e-invariant subclass of solutions, {\it i.e.}\ those with vanishing
total energy-momentum. Such a restriction is physically unwarranted;
instead, one treats the Poincar\'e group as a rigid symmetry that {\it
classifies} inequivalent solutions into multiplets according to their
Poincar\'e eigenvalues. 

     Without taking a firm position on the interpretation of duality
transformations that act nontrivially on the supergravity moduli, we would
nonetheless argue that the $U$-duality-symmetry little group should be
interpreted similarly to the asymptotic symmetry group in general
relativity. The asymptotic values of the scalar fields taken together with
the asymptotic values of the metric and of the antisymmetric-tensor gauge
fields constitute the full set of moduli of the theory. 

     Classically, the manifold in which the supergravity scalar fields take
their values is a coset space $G/H$, where $G$ is the relevant Cremmer-Julia
symmetry group $E_{r(+r)}$ and $H$ is its linearly-realised subgroup,
generally the automorphism group of the supersymmetry algebra. At this
classical level, the little group of the scalar moduli is simply $H$. When
one takes into account the quantum Dirac-Schwinger-Zwanziger charge
quantisation condition, the electric and magnetic charges of $p$-branes are
restricted to lie on a charge lattice. The subgroup of $G$ that respects
this charge quantisation is $G(\Z)$ \cite{ht}. In the context of string
theory, the $U$-duality symmetry $G(\Z)$ is interpreted as a gauge symmetry,
identifying vacua differing by $G(\Z)$ transformations so that the scalar
fields take their values in the manifold $G(\Z)\backslash G/H$. In the
special case of vanishing scalar moduli, the embedding of $G(\Z)$ into $G$
is particularly simple. Moreover, $G(\Z)$ then has a non-trivial
intersection $G(\Z)\cap H$ with the little group $H$ of the scalar moduli;
this intersection is in fact the Weyl group of the Cremmer-Julia symmetry
$G$ \cite{weyl}. 

     When one moves away on $G(\Z)\backslash G/H$ from the special point
with vanishing scalar moduli, both the embedding of $G(\Z)$ within $G$ and
the embedding of $H$ within $G$ change, in both cases by conjugations
involving the $G$ transformation moving the moduli to the new point. These
changes of embedding by conjugation move the two subgroups oppositely,
however.  Consequently, the two groups do not ``track''
together as one moves away from the special point on the scalar modulus
space, and so $G(\Z)\cap H$ becomes smaller than the Weyl group of $G$. In
fact, for generic scalar moduli on $G(\Z)\backslash G/H$, this intersection
will be trivial, {\it i.e.}\ the identity element alone. Consequently, the
Weyl group of $G$ is the {\it maximal} $U$-duality little group, achieved at
the special point of vanishing scalar moduli. 

\section{Modular-invariant 1-form solutions} 

    In the above discussion, we have not advocated treating the $U$-duality
group $G(\Z)$ as an {\it active} gauge symmetry, which would require {\it
invariance} under $G(\Z)$. To do so would for example rule out the standard
$p$-brane solutions, since they form non-trivial Weyl-group multiplets.
Nonetheless, it is interesting to enquire whether any solutions would
survive if one were to impose the restriction of full active $U$-duality
invariance. 

     We shall find that one may indeed construct a special class of
duality-invariant solutions, once again in the context of $(D-3)$ branes.
This class will be invariant under an $SL(2,\Z)$ duality symmetry, which in
$D=10$ type IIB supergravity is the full $G(\Z)$ duality group. In \cite{v},
a 7-brane solution \cite{ggp} of this kind was used as a background for
compactification of the type IIB string to a theory in $D=8$ that is dual to
the heterotic string compactified on $T^2$. (The fact that this solution
leads to a compactification of the transverse dimensions is again a
consequence of the conical asymptotic structure of transverse spacetime for
$(D-3)$ branes.) Since the 7-brane solution is invariant under the
$SL(2,\Z)$ symmetry, the transverse compactification preserves this
symmetry, and this fact is crucial for establishing a duality relation with
the heterotic string compactified on a torus. The 7-brane solution in $D=10$
type IIB supergravity preserves $\ft12$ of the supersymmetry, corresponding
to the Lagrangian (\ref{genlag}) with $\Delta =4$. The form of this solution
is very different from the usual $(D-3)$-brane soliton, and exploits special
features of the two-dimensional transverse space, which may be viewed as a
complex plane on which modular-invariant functions can be defined. As with
the usual $(D-3)$-brane solution discussed above, the modular-invariant
solution allows only a finite number of centers before the transverse space
becomes compact. In fact for regularity of such a compactified solution,
there must be exactly 24 centers \cite{gsvy}. 

      In lower dimensional supergravities, where the number of 1-form field
strengths proliferates, we can have supersymmetric solutions involving $k$
field strengths, giving rise to $\Delta =4/k$.  In this section, we shall
show that one can in fact construct a $p$-brane solution that is $SL(2,\R)$
invariant for any value of $\Delta$.  We shall then show that such a
solution must have exactly $6\Delta$ centers in order to obtain a regular
compactification of the transverse space.  The relevant part of the
Lagrangian, describing the metric, dilaton and 1-form field strength in $D$
dimensions is given by (\ref{genlag}) with $n=1$.  The Lagrangian can be
cast into the form 
\be
e^{-1} {\cal L} = R - \fft{2}{a^2} \fft{|\del \tau|^2}{\tau_2^2}\ ,
\label{lag1f}
\ee
where $\tau=\tau_1 + i \tau_2= \ft{a}2 \chi + ie^{-\ft{a}{2} \phi}$, and the 
$\chi$ is the 0-form potential for the 1-form field strength.  The dilaton 
prefactor for this field strength has $a^2= \Delta$, since $\td d=0$.  Thus
when $\Delta=4$, the above Lagrangian coincides with the one discussed in
\cite{gsvy,ggp,v}.  The Lagrangian (\ref{lag1f}) has an $SL(2,\R)$ symmetry
\be
\tau \longrightarrow \tau'=\fft{a \tau + b}{c \tau + d}\ ,\label{sl2r}
\ee
where $ad - bc =1$. The ansatz for the metric is given by $ds^2= dx^\mu
dx^\nu\eta_{\mu\nu} + e^{2B} dzd\bar z$, where $B$ is a function of $z=y_1 + 
{\rm i} y_2$ and $\bar z= y_1 - {\rm i} y_2$.  The equation of motion for
$\tau$ is given by 
\be
\del\bar \del \tau + \fft{2\del\tau\bar \del \tau}{\bar \tau - \tau}=0\ .
\label{eom1f1}
\ee
Following \cite{gsvy}, we make the holomorphic ansatz $\bar \del \tau= 0$, and
hence (\ref{eom1f1}) implies that $\del\bar\del \tau =0$.  The metric
equation, after imposing this ansatz, becomes 
\be
\del\bar \del B = \ft{2}{\Delta}\del\bar\del \log \tau_2\ ,\label{eom1f2}
\ee
The usual type of 1-form $p$-brane solutions discussed in the previous section
correspond to  taking $\tau = i\lambda\log z$ and $B= (2/\Delta) \log
\tau_2=-\phi/\sqrt\Delta$.  Such solutions are obviously not invariant under
$SL(2,\R)$, since $B$ is in that case a function of $\tau_2$ only.  In
\cite{gsvy}, an $SL(2,\R)$-invariant solution was constructed for the case
$\Delta=4$.  For other values of $\Delta$, it follows from (\ref{eom1f1}) and
(\ref{eom1f2}) that the solution for $\tau$ is unchanged, while the solution
for $B$ is simply rescaled by a factor of $4/\Delta$.  Thus following
\cite{gsvy}, the asymptotic behaviour of $B$ at large $|z|$ is given by 
\be
e^{2B} \sim (z\bar z)^{-N/(3\Delta)}\ .
\ee
The exact invariant solution is given by
\be
e^{2B} =\tau_2 \eta^2 \bar \eta^2 \left| \prod_{i=1}^N (z-z_i)^{-1/(3\Delta)}
\right|^2\ ,\label{mis}
\ee
where $\eta = e^{i\pi \tau/12} \prod_n (1 - e^{i 2n\pi \tau})$ is the 
Dedekind $\eta$ function.  It was shown in \cite{gsvy}, for the case
$\Delta=4$,  that each center contributes a deficit angle $\ft{\pi}{6}$ in the
transverse  space, and thus it remains non-compact if $N\le 11$.  In order to
be regular  and compact, one must take $N=24$ \cite{gsvy}.  We can therefore
immediately  generalise these results to the other values of $\Delta$.  In
particular,  there must be $N=6\Delta$ centers in order to achieve a regular
compactified  transverse space.

      We shall now discuss the supersymmetry of these modular-invariant 
solutions.  First we consider the supersymmetry of the 7-brane in type 
IIB supergravity in $D=10$, where (\ref{lag1f}) with $a^2 =4$ is the full 
scalar Lagrangian of the theory.  In a bosonic background where only $\tau$ 
and $g_{\sst{MN}}$ are excited, the supersymmetry transformation rules for
the complex spin $\ft12$ and spin $\ft32$ fermions are \cite{s}
\be
\delta \lambda = \fft{\tau^* -i}{\tau_2(\tau + i)} \Gamma^{\sst M}
\del_{\sst M} \tau \epsilon^*\ ,\qquad
\delta \psi_{\sst M} =\del_{\sst{M}}\epsilon +
\ft14\omega_{\sst M}^{\sst{AB}} \Gamma_{\sst{AB}} \epsilon
-\ft{\rm i}{2} Q_{\sst M}\epsilon\ ,\label{susu1}
\ee
where $\epsilon=\epsilon_1+ i\epsilon_2$, $\epsilon^*=\epsilon_1 -i 
\epsilon_2$, and the composite $U(1)$ gauge potential is given by
\be
Q_\sst{M} = -{1\over 4 \tau_2}\Big\{{\tau-{\rm i}\over \tau^*-{\rm i}} \,
\del_{\sst M} \tau^* + {\tau^* +{\rm i} \over \tau+{\rm i}}\, \del_{\sst M}
\tau \Big\}\ .\label{Qexp} 
\ee
Note that $Q_{\sst M}$ is invariant under the $SL(2,\R)$ transformation 
(\ref{sl2r}), modulo a compensating local $U(1)$ gauge transformation.   

     The only non-vanishing spin connection, in the 7-brane metric $ds^2=
dx^\mu dx^\nu \eta_{\mu\nu} + e^{2B} dy^m dy^m$, is given by $\omega^{12}_z=
{\rm i} \del B$, $\omega^{12}_{\bar z}= -\bar\del B$, where 1 and 2 refer to
flat indices in the two-dimensional transverse space.  We see that if
$\epsilon$ is chosen so that $\Gamma^1\epsilon={\rm i} \Gamma^2\epsilon$,
then $\Gamma^z \epsilon^*=0$, and hence $\delta\lambda=0$ by virtue of the
holomorphicity condition $\bar\del \tau=0$.   For the variation
$\delta\psi_{\sst M}$ to vanish also, we must have $\omega^{12}_{\sst M}=
Q_{\sst M}$, up to a local $U(1)$ gauge transformation.  Thus we must have
${\rm i}\del B =\widetilde Q_z\equiv Q_z + \del f(\tau)$, where $f$ is a
holomorphic function that must be added to $Q_z$ in order to make
$\widetilde Q_z$ $SL(2,\R)$ invariant. The function $f$ has to satisfy 
\be
\del f(\tau') -\del f(\tau) = Q_z(\tau) - Q_z(\tau') =
-\del [(2{\rm i} \log \Big((\tau + i) (\tau + \ft{b+{\rm i}d}{a + {\rm i}c})
\Big)]\ .
\ee
One way to solve this is to take $f(z) = -{\rm i} \log [\del(\tau +{\rm
i})^3]$.  

     It is easy to verify that the condition ${\rm i}\del B = \td Q_z$ that
is required for supersymmetry yields (\ref{eom1f2}) after acting on it with
$\bar\del$.  Thus this supersymmetry condition is the $SL(2,\R)$-invariant
first integral of the equation of motion for the metric.  The
$SL(2,\R)$-invariant expression (\ref{mis}) (with $\Delta=4$) for the 7-brane
is a solution of this first integral. 

      As we mentioned above, the 1-form field strength in $D=10$ type IIB
supergravity can also give rise to a 7-brane of the usual $SL(2,\R)$
non-invariant kind, with $\tau={\rm i}\lambda \log z$ and $B=\ft12
\log\tau_2$.  However, this solution is not supersymmetric since, as we saw
above, supersymmetry requires not only that $\tau$ be a holomorphic function
but also that $B$ be $SL(2,\R)$ invariant (up to a $U(1)$ gauge
transformation).  In $D\le9$, on the other hand, the $SL(2,\R)$ non-invariant
$(D-3)$-brane solutions can be supersymmetric \cite{stainless}. The first
example is the isotropic 6-brane in $D=9$ \cite{stainless}, which preserves
$\ft12$ of the supersymmetry.  Since this solution is not $SL(2,\R)$
invariant, it is not related to the above supersymmetric 7-brane in type IIB
supergravity by dimensional oxidation.  It was accordingly called a
stainless 6-brane in \cite{stainless}.  On the other hand, double
dimensional reduction of the 7-brane in type IIB supergravity will give rise
to their own trajectory of supersymmetric $SL(2,\R)$ invariant
$(D-3)$-branes in lower dimensions. This $SL(2,\R)$ symmetry is the full
symmetry group in type IIB supergravity.  In lower dimensions, the symmetry
group of the supergravity theories enlarges, and so these solutions  are
invariant only under a subgroup of the full symmetry group. The question
naturally arises whether one can construct further $(D-3)$-brane  solutions
that are invariant under the larger symmetries. 

     In lower-dimensional supergravities, the number of axions proliferates, 
and $(D-3)$-branes can be constructed using multiple field strengths. The
relevant part of the Lagrangian can be reduced to (\ref{lag1f}) in a manner
that is consistent with the equations of motion from the full Lagrangian
\cite{lp}. In particular, we can construct the usual supersymmetric solutions
with up to $k=7$ 1-form field strengths in $D=4$, and up to $k=8$ in
$D=3$ \cite{lp}.  The values of $\Delta$ for these supersymmetric
solutions are given by $\Delta  = 4/k$.  We have seen in this section that
such field configurations can also be used to construct $SL(2,\R)$-invariant
solutions.  The supersymmetry of  these solutions requires further
investigation; however, we expect that  they preserve the same fraction of the
supersymmetry as their $SL(2,\R)$ non-invariant counterparts.

\section*{Acknowledgements}

     We are grateful to P.K.\ Townsend for stimulating discussions of these
issues. K.S.S. would like to thank the Departments of Physics at Texas A\&M
University and at KU Leuven for hospitality at different periods during the
course of the work.

\section*{Note added}

       After this paper was circulated, we learned that the solution for
four intersecting membranes was also constructed in \cite{gkt}.  The 
vertical dimensional reduction of $(D-3)$-branes,  and the relation to 
massive supergravities, has been studied recently in \cite{clpst}.

\end{document}